\documentstyle[11pt,mrs2001,epsfig1_7a]{article}
\begin{document}
\title{TOPOLOGICAL LENSING: IS THE MATTER BOTH HERE AND THERE?}

\author{B.F. ROUKEMA$^{1,2,3}$, A. MARECKI$^4$, S. BAJTLIK$^2$, 
R.E. SPENCER$^5$}
\affil{$^1$DARC/LUTH, Observatoire de Paris-Meudon, 5, place Jules Janssen, 
F-92.195, Meudon Cedex, France}
\affil{$^2$Nicolaus Copernicus Astronomical Center, 
ul. Bartycka 18, 00-716 Warsaw, Poland}
\affil{$^3$University of Warsaw, Krakowskie Przedmie\'scie 26/28, 
00-927 Warsaw, Poland}
\affil{$^4$Toru\'n Centre for Astronomy, Nicolas Copernicus University,
ul. Gagarina 11, PL-87-100 Toru\'n, Poland}
\affil{$^5$Jodrell Bank Observatory, Macclesfield, 
Cheshire SK11 9DL, United Kingdom}

\newcommand\Omm{\Omega_{\mbox{\rm \small m}}}
\newcommand\hMpc{\mbox{$\,h^{-1}\,$Mpc}}
\newcommand\hGpc{\mbox{$h^{-1}$ Gpc}}
\newcommand\hkpc{\mbox{$h^{-1}$ kpc}}
\newcommand\hGyr{\mbox{$h^{-1}$ Gyr}}
\newcommand\hMyr{\mbox{$h^{-1}$ Myr}}

\newcommand\ltapprox{\,\lower.6ex\hbox{$\buildrel <\over \sim$} \, }

\begin{abstract}
   If the Universe satisfies a perturbed Friedmann-Lema\^{\i}tre model,
then the bright matter (e.g. radio-loud active galactic nuclei,
RLAGNs) may be topologically lensed by global geometry. The generation
of candidate topological lensing pairs of RLAGNs, which may be double
images of single objects seen at very different celestial positions,
provides very easily falsifiable (given moderate telescope time)
cosmological geometry hypotheses.
\end{abstract}

If the matter is both here and there, i.e. if it is topologically
lensed, then it would
be good that we know about this. 

%\section{Introduction}

In order to measure the global geometry of the Universe, according to the
perturbed Friedmann-Lema\^{\i}tre model, it is necessary to measure
the topology of the Universe 
\cite{LaLu95,LR99,Stark98,BR99,Rouk2001MG9}.
Previous work on searches for multiple imaging of quasars considered
these as point objects. 
In contrast, a new strategy is 
to search for radio-loud active galactic nuclei (RLAGNs)
at nearly identical redshifts. This makes
it possible to obtain pairs of 
moderate to high redshift ($0.5 \ltapprox z \ltapprox 4$) 
candidate topologically lensed
objects for which high resolution imaging exists or is readily
obtainable. This
implies that (i) the angular orientations of the radio lobes 
imply geometrical constraints in addition to the objects' locations
in comoving space; and (ii) consistency of the sizes of the 
images based on 
reasonable (i.e. positive and not too slow) expansion speeds
provide another constraint on the lensing hypothesis.

Here, just one candidate topologically lensed pair of images of RLAGNs 
is presented. The images of the RLAGNs 3C186 and 4C+36.21 are surprisingly
similar. The redshift of 3C186 is known: $z=1.063$. 

The redshift of 4C+36.21 is unknown (we intend to measure this redshift).

Could these be two topological images of a single RLAGN, separated by
just a short time delay? 

Observation will provide a very strong
test of the hypothesis that this is a case of topological lensing.

4C+36.21 is seen with a linear size in
proper units of about 1.6{\hkpc}, while 3C186 is 6.5{\hkpc} in size,
implying that the 4C+36.21 image must be just slightly
earlier in time than 3C186, but not too much earlier.

Unless the measured redshift of 4C+36.21 is found to lie in the
very narrow range $1.0630< z \ltapprox 1.0635$, then identity would
require either 
a jet which contracts or 
a jet which expands much slower than $0.01c$,
neither of which are physically reasonable.

% %\def\tMSpen{
% \begin{table*}
% \caption{The pair of very compact steep spectrum sources (3C186,4C+36.21), 
% which would need to be at nearly identical redshifts in order to qualify
% as a candidate topological image pair. Listed are the names, 
% redshifts $z_i$ (where ``?'' indicates that 
% the redshift of 4C+36.21 is at present unmeasured), interval in redshift 
% $\Delta z$, interval in time $\Delta t \equiv t(z_2) - t(z_1)$ 
% for density parameter and cosmological constant 
% $(\Omm=0.3,\Omega_\Lambda=0.7)$, 
% linear sizes $d_i$ in proper units (estimated as half the
% angle between points of maximum flux in the two lobes in the 20cm images), 
% and ``na\"{\i}ve'' apparent sky-plane expansion speed $v$.
% \label{t-MSpen}}
% $$\begin{array}{cc cc c ccccc} \hline 
% \rule{0ex}{2.5ex}
% \mbox{RLAGN}_1 & z_1 &
% \mbox{RLAGN}_2 & z_2 & \Delta z &
% \Delta t \mbox{({\hMyr})} &
% d_1 \mbox{({\hkpc})} &d_2 \mbox{({\hkpc})} & v/c \\
% 4C+36.2 &  1.0635?& 3C186   &  1.0630&  0.001&   1.3&    1.6&    6.5&  0.012 \\
% 4C+36.2 &  1.0631?& 3C186   &  1.0630&  0.000&   0.3&    1.6&    6.5&  0.061 \\
% \hline
% \end{array}$$
% \end{table*}
% %}  %\def\tMSpen
% 

\begin{figure}
\plotone{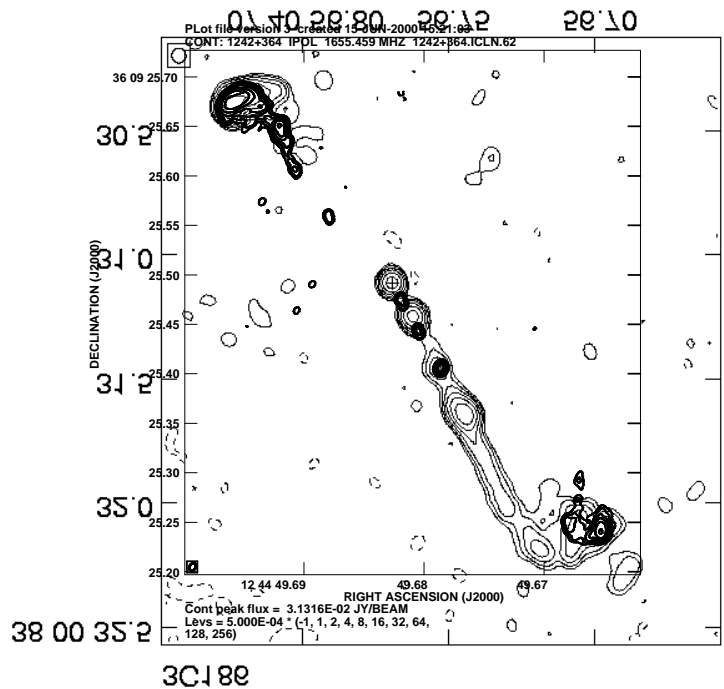}{.6\textwidth}
\caption{The two steep compact spectrum sources. The image 
of 4C+36.21 (at 18cm) is shown in heavy contours; the image of 3C186
(at 6cm) is shown in light contours and reflected North-South. 
Either (i) the two RLAGN images are of physically distinct objects
which display a similar physical process or (ii) they are two topologically
lensed images of
a single object.
}
\end{figure}

\acknowledgements{
This research has been supported by the 
Polish Council for Scientific Research Grant
%%ancien KBN 2 P03D 008 13 
KBN 2 P03D 017 19
and has benefited from 
the Programme jumelage 16 astronomie 
France/Pologne (CNRS/PAN) of the Minist\`ere de la recherche et
de la technologie (France).}

\newcommand\joref[5]{#1, #5, {#2 }{#3, } #4}  % journal reference
%   author, journal, volume, page, year
\newcommand\epref[3]{#1, #3, #2}
% e-print reference (http://xxx.lanl.gov or mirror http://babbage.sissa.it )
% author, archive-subj/0111052, year
\newcommand\cqg{ClassQuantGrav}

\vfill
\end{document}